\documentclass[preprint2]{aastex}
\shortauthors{R. Kehoe et al.}
\shorttitle{Short and Long GRB Counterpart Search}

\begin{document}

\title{A Search for Early Optical Emission from Short and Long Duration Gamma-ray
Bursts}

\author{Robert Kehoe$^{1,4}$, Carl Akerlof$^1$, Richard Balsano$^2$, 
Scott Barthelmy$^5$, 
Jeff Bloch$^2$, Paul Butterworth$^5$, Don Casperson$^2$, 
Tom Cline$^5$, Sandra Fletcher$^2$, 
Galen Gisler$^2$, Kevin Hurley$^7$, Marc Kippen$^6$, 
Brian Lee$^{1,8}$, Stuart Marshall$^3$, Tim McKay$^1$, 
Eli Rykoff$^1$, Don Smith$^1$,
Tom Vestrand$^2$ and Jim Wren$^2$}

\affil{$^1$University of Michigan, Ann Arbor, MI 48109}
\affil{$^2$Los Alamos National Laboratory, Los Alamos, NM 87545}
\affil{$^3$Lawrence Livermore National Laboratory, Livermore, CA 94550}
\affil{$^4$Michigan State University, East Lansing, MI 48824}
\affil{$^5$NASA/Goddard Space Flight Center, Greenbelt, MD 20771}
\affil{$^6$NASA/Marshall Space Flight Center, Huntsville, AL 35805}
\affil{$^7$Space Sciences Laboratory, University of California, Berkeley, 
CA 94720-7450}
\affil{$^8$Fermi National Accelerator Laboratory, Batavia, IL 60510}

\begin{abstract}
Gamma-ray bursts of short duration may harbor vital clues to the range
of phenomena producing bursts. However, recent progress from the 
observation of optical counterparts has not benefitted the study of
short bursts.  We have searched for early
optical emission from six gamma-ray bursts using the ROTSE-I telephoto array.
Three of these events were of short duration, including GRB 980527 which is
among the brightest short bursts yet observed. The data consist of unfiltered
CCD optical images taken in response to BATSE triggers delivered via the GCN.
For the first time, we have analyzed the entire \( 16^\circ\times
16^\circ \) field covered for five of these bursts. In addition, we discuss a search for
the optical counterpart to GRB 000201, a well-localized long burst. Single
image sensitivities range from 13th to 14th magnitude around 10 s after the
initial burst detection, and 14 - 15.8 one hour later. No new
optical counterparts were discovered in this analysis suggesting
short burst optical and gamma-ray fluxes are uncorrelated.
\end{abstract}

\keywords{gamma rays: bursts, observations}

\clearpage

\section{Introduction}

During the last decade, gamma-ray burst classification has emerged
as a promising tool to understand these events. In the pre-BATSE data, there
were indications of a bimodality in the temporal durations of non-repeating
bursts \cite{hurley92}, as well as a modest correlation of duration with spectral
hardness \cite{dezalay92}. These results were confirmed by data from the Burst and
Transient Source Experiment (BATSE) which show that short bursts inhabit a distinct
region of the duration/spectral hardness parameter space
\cite{kouvelioutou93}.  More recently, theoretical
work has made use of the BATSE archive to attempt to determine the processes
which distinguish short and long bursts. There is some evidence that short
GRBs arise from internal shock processes \cite{nakar}, and that they are 
differentiated from long bursts by the ejecta shell geometry
(eg. \cite{kobayashi}).  Alternatively,
these two classes may arise from different accretion disk states
around progenitor
black holes \cite{vanputten}. Indications that very short bursts exhibit an unusually
uniform spectral hardness distribution may hint at yet a third class of GRB
\cite{cline99}.

Despite what has been gleaned from the gamma-ray emission, the study of short bursts
has not been benefited by the more recent advances brought about by optical
afterglow (eg. \cite{metzger97}, \cite{kulkarni98}) and burst \cite{nature} observations, all
of which have identified only long-duration bursts. One of the main 
reasons for this bias stem from difficulty in obtaining well-localized
final positions for these events.
To date, the main source of well-localized burst notices for late optical study
has been the BeppoSAX satellite \cite{feroci97}, but the 1 s integration time of the GRBM prevents
efficient detection of short bursts. In addition, prompt wide-field searches
such as ROTSE-I and LOTIS \cite{lotis} have been hampered by the practical difficulty
of searching their \~{}250 sq. degree fields. At a sensitivity of 15th magnitude,
there are of order \( 10^{5} \) stars in these fields, and approximately a
few million photometric measurements in a typical ROTSE-I trigger response.
Burst finding necessitates well-controlled photometry across an entire field,
as well as a detailed knowledge of the hardware behavior (eg. bad pixels, position
resolution). Although this limitation was overcome for long bursts by using
localizations from the Inter-Planetary Network (IPN, \cite{hurley99}) to reduce the
search area \cite{rotse00}, no such localizations were available for the short bursts
on which ROTSE-I triggered.

To overcome this limitation, we have designed an analysis specifically
to confidently analyze many observations of a large field. This paper presents
the results of a search for optical counterparts to five bursts in the ROTSE-I
trigger data sample using these techniques. An additional burst, studied with
our previous method, is also presented.

\section{Observations and Reduction}

The ROTSE-I array used in these observations consists of four telephoto lenses
comounted in a \( 2\times 2 \) geometry on a lightweight platform which allows
rapid slews and excellent pointing accuracy relative to the \( 16.4^{o}\times 16.4^{o} \)
total field-of-view (14.4'' pixel scale, see \cite{kehoe00a} for more details). From
March, 1998 until the deorbiting of the Compton Gamma-Ray Observatory (CGRO)
in June 2000, 57 GRB triggers were received via the GRB Coordinates
Network (GCN, \cite{gcn}, \cite{bacodine}) derived from BATSE data
\cite{paciesas}. To date, we have successfully analyzed data for a subset of seven
long duration bursts possessing small localization errors, yielding one detected
prompt optical counterpart \cite{nature} and six non-detections \cite{rotse00}. This
paper discusses a search in a set of six additional triggers taken during the
full period of BATSE activity while ROTSE-I was automated. Five of these triggers
were selected for this analysis because (1) they fall into the short duration
class of bursts, (2) their most probable known position is within \( 5^{o} \)
of the ROTSE-I pointing, or (3) we have a thorough, photometric-quality set
of images from most of the cameras. The sixth burst, GRB 000201, is included
because the analysis of the known IPN diamond localizing this burst was straightforward,
and other optical follow-up has been performed \cite{boer}. Preliminary
results for GRB 980527 and GRB 000201 have been previously presented in \cite{kehoe00a,kehoe00b},
respectively. This paper supersedes and improves upon those results. Some important
burst parameters from the BATSE observations, as well as the ROTSE-I spatial
and temporal coverages are itemized in Table 1. The spatial coverage is
defined as the probability that the burst location was imaged in our search,
and it is based on the statistical and systematic uncertainties in the best
BATSE or IPN localizations.

The observations scheduled in response to a trigger vary somewhat depending
on trigger type and delay from the time of the burst. The three trigger types
involved in this analysis are `Original' which arrives around 7 seconds after
the burst start, `Final' which arrives approximately 1 minute later,
and `MAXBC' which arrives about 5 to 10 minutes after the burst.  For
all of the trigger responses taken with ROTSE-I, 5 s exposures are
taken during the first minute of the burst.
Prior to December 1998, the exposures were lengthened to 25 s and then 125 s
as the delay progressed from 1 minute to approximately 10
minutes. After January 1999, these longer exposure lengths were
reduced to 20 s and 80 s, respectively.
There are occasional gaps in temporal coverage for the less accurate 
`Original' and `MAXBC' triggers because we hedged our bets and
scheduled tiling sequences during the burst response.

GRB 980527 was one of the first well-localized bursts observed by
ROTSE-I.  This is a very intense, spectrally hard burst lasting 
approximately 0.1 s. The first
images were taken within 12 s of the burst start and consist of two groups of
five exposures of 5 s and 25 s length with a tiling gap in between. After this,
five 25 s exposures were taken centered on the subsequent MAXBC localization.
Unfortunately, the MAXBC localization is more distant from the most likely position
for this burst, reducing our coverage at later times. In addition, camera `a'
was not functioning which reduces our coverage to 86\% (stat.+sys.). The only
trigger localizations provided for both GRB 990323 and GRB 990808 were of the
'MAXBC' variety, and so our imaging begins more than 10 minutes after these
bursts. Both GRB 991028 and GRB 991228 possess `Original' and `Final'
localizations.  GRB 991028 is a spectrally hard short burst and these
on-line positions are a good match to that determined off-line by the
BATSE team.  Hardware behavior suffered in two ways
for the GRB 991228 trigger response: camera `c' did not take useable data, and
camera `d' exhibited a mild charge-transfer problem. For GRB 000201, we only
triggered on the `Final' localization since the `Original' position was deemed
unobservable by the automated scheduler. The best localization for
this burst consists of an
IPN diamond derived from BATSE, ULYSSES and NEAR data, and is well-covered by
the ROTSE-I images. Observing conditions for all of these bursts were good. 

The reduction of the data proceeds as in \cite{kehoe00a}. Raw images are dark subtracted
and flat-fielded by using darks and sky-flats generated on the night of the
trigger. Clustering of the corrected image is performed with SExtractor \cite{sextr}
utilizing a background mesh segmented by 32 pixel increments. Raw magnitudes
are based on \( 5\times 5 \) pixel apertures. Astrometric calibration and determination
of the overall zero-point for the image are performed by comparison to the Hipparchos
catalog \cite{tycho}. During this step, a bad pixel template, derived from the
raw darks for each camera, is used to flag objects containing such pixels within
their apertures. A systematic error is assigned to the source based on the observed
fluctuations of the constituent bad pixels. 

Individual calibrated source lists are matched up to one another to form preliminary
lightcurves. Any observation of an object is tagged as bad in which the position lies over
one half pixel from the source's mean location. Final photometric calibration
across the image is then performed by determining the median magnitude for a
set of template sources for each 100 pixel \( \times  \) 100 pixel image subregion,
and using these to derive a map of offsets and offset variances in each subregion.
All objects are then corrected by a bilinear interpolation of this relative
photometry map, and a systematic error is assigned based on the offset 
variance.

\section{Analysis and Discussion}

The analysis of the burst fields proceeded on two paths: a small field search
applied to GRB 000201, and a wide field search applied to the other five. In
both, candidate objects are rejected if they do not occur in at least
one consecutive pair of observations. In addition, an object's individual observations
are ignored if they are saturated or at an image edge. From here, the two analyses
differ.

Because the IPN localization for GRB 000201 dramatically reduced the background
for a counterpart search, we employed a loose lightcurve selection. During the
observations in which a candidate is detected, there must be at least
one pair of good
measurements which exhibits a variation in excess of \( 0.5+5\sigma  \), where
\( \sigma  \) is the sum in quadrature of the statistical and systematic errors
of the two observations involved. Because it was the best localization early,
we searched within the 1 \( \sigma  \) (stat.) limits of the later BATSE
LocBurst position. We have also checked the location of a purported
optical counterpart detection \cite{boer}.
No counterparts were identified. An initial IPN localization became available
on Feb. 9, and we searched within a surrounding box having sides with \( \alpha =138.75^{o}-139.25^{o} \)
and \( \delta =17.95^{o}-18.45^{o} \). The final IPN diamond which was obtained
later comprises the inner 20\% of this region. We found no
counterparts in this region. 

This IPN-based analysis gives insufficient control of backgrounds in wide-field
searches. To address this, we first attempted an image differencing approach
on GRB 980527 \cite{kehoe00a}. This method's effectiveness
was reduced, however, because blending is not a
critical problem in ROTSE images away from the galactic plane, and the typical
PSF is very undersampled. In addition, it removes the diagnostic information
present in the original image with which we can either correct occasional mild
photometric variations, or flag problem observations. As a result, we perform
our wide-field search by extending our IPN-based analysis with a more 
restrictive lightcurve variation selection, as well as a more
stringent requirement on local
observation quality. Individual observations are rejected if: a bad pixel lies
within the aperture, there is a large offset from the mean position (ie.$> 0.5$
pixel), photometry in the local image subregion has a large ($> 0.1$ mag) standard
deviation, or there are $< 5$ template stars per local subregion. Lightcurve
cuts were chosen based on a comparison of simulated bursts with
power-law lightcurves vs. typical backgrounds observed in ROTSE trigger data. 
We require at least one variation passing a logical AND of $> 0.5$
mag. and $> 5\sigma$ (stat.+sys.).
In addition, we require that the overall lightcurve
(using good observations only) differs from constant by a 
\( \chi _{cl}^{2}>3.0 \)
per degree-of-freedom, where the largest variation is excluded from the calculation.
This last criterion also implies that the source is detected in 3 good, not
necessarily consecutive, observations.

No candidates were observed for GRB 990323, GRB 990808 or GRB 991028
after these selections. A few objects passed the selection for GRB 980527,
but most of these correspond to known sources in the USNO A-2.0 catalog \cite{usno}.
Examination of the images for the remaining candidates revealed them to
be incompletely removed bad pixels. For GRB 991228, the source of the candidates
was solely due to a charge-transfer problem giving rise to spurious transients
in the trailing charge. The limits for these bursts are shown in Figure 1, and
are itemized in Table 2 for up to three discrete epochs.

All three short bursts have very prompt trigger responses, and two, GRB 980527
and GRB 991028, provide a probable coverage of the allowed error region. The
earliest limits from these two bursts are 13.1 and 13.8 mag at 9.9 s and 12.2
s, respectively. Later limits range from magnitude 14 to 15. Our non-detections
for these bursts, which have very different durations, suggest that
short bursts do not usually give rise to optical
emission brighter than 13th or 14th magnitude in the first few minutes after
the burst. Unfortunately, we do not have data for the first 10 seconds
of the burst which may be a critical phase for short duration events.

Among the three long duration bursts analyzed, two have a high coverage probability,
and one (GRB 000201) has images starting soon after the burst. The best limits
for this sample are 14.0 mag at 95.9 s and 15.8 mag late. The non-detection
of a counterpart in these cases reinforces our previous conclusion that early
optical emission from GRBs is not typically brighter than 14th magnitude, and
it is fainter than 16th magnitude around 1 hour after the burst.

In an effort to compare this data to the only burst in which prompt optical
emission has been observed, we have replotted the limits in Figure 2
after adjusting
for their peak flux in 64 ms binning of the 50-300 keV BATSE data, as compared
to GRB 990123. An optical burst from GRB 980527, GRB
981028 and GRB 000201 would have been observable if optical flux and 
gamma-ray flux were highly correlated.  

More prompt optical data are clearly needed.  This analysis, which is the
first attempt in this direction, signals a hopeful future for more
sensitive work.
The methods described here are directly applicable to the ROTSE-III
imaging which will be the mainstay of this effort in the future.
It should now be possible to perform IPN-based searches to much deeper 
magnitudes than with ROTSE-I. More importantly for early optical
study, we are now implementing
an improved pipeline to find an optical transient in a fully automated
and rapid way based on HETE-2 triggers.

\begin{acknowledgments}
        We thank the BATSE team for their GRB data.  We also thank the 
NEAR and ULYSSES teams for their GRB 000201 data.  ROTSE is supported by NASA 
under $SR\&T$ grant NAG5-5101, the NSF under grants AST-9703282 and 
AST-9970818, the Research Corporation, the University of Michigan, and the 
Planetary Society.  Work performed at LANL is supported by the DOE under
contract W-7405-ENG-36.  Work performed at LLNL is supported by the DOE
under contract W-7405-ENG-48.  Kevin Hurley is grateful for Ulysses
support under JPL Contract 958056, and for NEAR support under NASA
grant NAG5-3500 and NAG5-9503.
\end{acknowledgments}

\clearpage

%\vspace{0.3cm}
%{\par\centering \resizebox*{3.5in}{!}{\includegraphics{norm.eps}} \par}
%\vspace{0.3cm}

\figcaption[norm.ps]{$m_{ROTSE}$ limiting magnitudes vs. time 
        after gamma-ray onset.  GRB 990123 optical burst detections 
        are shown for comparison.  \label{fig:grb_timehist}}

%\vspace{0.3cm}
%{\par\centering \resizebox*{3.5in}{!}{\includegraphics{flux.eps}} \par}
%\vspace{0.3cm}

\figcaption[flux.ps]{Flux rescaled limits for six GRBs vs. 
        time after gamma-ray onset.  If optical emission were positively 
        correlated with peak gamma-ray flux, ROTSE-I would have detected 
        optical bursts for GRB 980527, GRB 991028, GRB 991228, and GRB 000201. 
        \label{fig:rflux}}

\clearpage

\begin{table}
  \begin{center}
  \begin{tabular}{cccccccc}
	GRB  & trigger & $T_{50}$ &$T_{90}$ &$\phi _{peak}$ & fluence& coverage&$t_{+}$\\
        \hline
	980527	& 6788 & 0.049	& 0.092	& 21.1	& 6.5	& 86(70)& 12.19\\
	990323	& 7489 & 85.0	& 118.8	& 0.51	& 28.2	& 20	& 640.62\\
	990808	& 7704 & 42.0	& 75.8	& 0.81	& 8.9	& 75	& 716.38\\
	991028	& 7827 & 0.5	& 2.3	& 1.2	& 0.6	& 70	& 9.90\\
	991228	& 7922 & 0.27	& 0.42	& 2.8	& 1.0	& 5	& 17.65\\
	000201	& 7976 & 48.3	& 95	& 3.3	& 145	& 100	& 85.89\\
  \end{tabular}
  \caption{Characteristics of six bursts responded to by ROTSE-I.
	Corresponding information for GRB990123 is given in \cite{7}).
	The columns specify: GRB date, BATSE trigger number, duration
	in seconds where endpoint fluxes are 50\% of peak, duration where
	endpoint fluxes are 10\% of peak, peak flux ($\phi_{peak}$, 
	in $photons/cm^2/s$), fluence ($\times 10^{-7} erg/cm^2$), 
        coverage of the GRB probability (\%), and start time ($t_+$, in sec.) for 
        first image recorded.  The peak flux is taken using 64 ms
	binning, except for GRB 990323 and GRB 990808 for which only
	1024 ms binning was available.  Coverages for epochs with pointings
	different than the first epoch are indicated in parentheses.
        \label{tab:sixgrbs}}
  \end{center}
\end{table}

\begin{table}
  \begin{center} 
  \begin{tabular}{cccccccccc}
        date & $t_1$ & $\Delta t_1$ & $m_{ROTSE}(t_1)$ & 
               $t_2$ & $\Delta t_2$ & $m_{ROTSE}(t_2)$ & 
               $t_3$ & $\Delta t_3$ & $m_{ROTSE}(t_3)$\\
        \hline

	980527 & 14.69 & 5  & 13.76 & 208.35 & 25 & 14.84 & 617.64  & 25 & 14.98\\
	990323 & -     & -  & -     & 650.62 & 20 & 14.80 & 1048.06 & 80 & 15.59\\
	990808 & -     & -  & -     & 726.38 & 20 & 14.48 & 1122.25 & 80 & 15.10\\
	991028 & 12.40 & 5  & 13.06 & 200.02 & 20 & 13.65 & 397.85  & 80 & 13.85\\
	991228 & 20.15 & 5  & 13.71 & 81.49  & 20 & 14.78 & 666.16  & 80 & 15.33\\
	000201 & 95.89 & 20 & 14.03 & 680.00 & 20 & 15.14 & 2871.64 & 365.6 & 15.76\\
  \end{tabular}
  \caption{Summary of limits for six bursts responded to by ROTSE-I.  Columns 
        list up to three epochs (middle of exposure, in sec.), and their
        exposure length (sec.) and sensitivity.
        \label{tab:sixlims}}
  \end{center} 
\end{table}

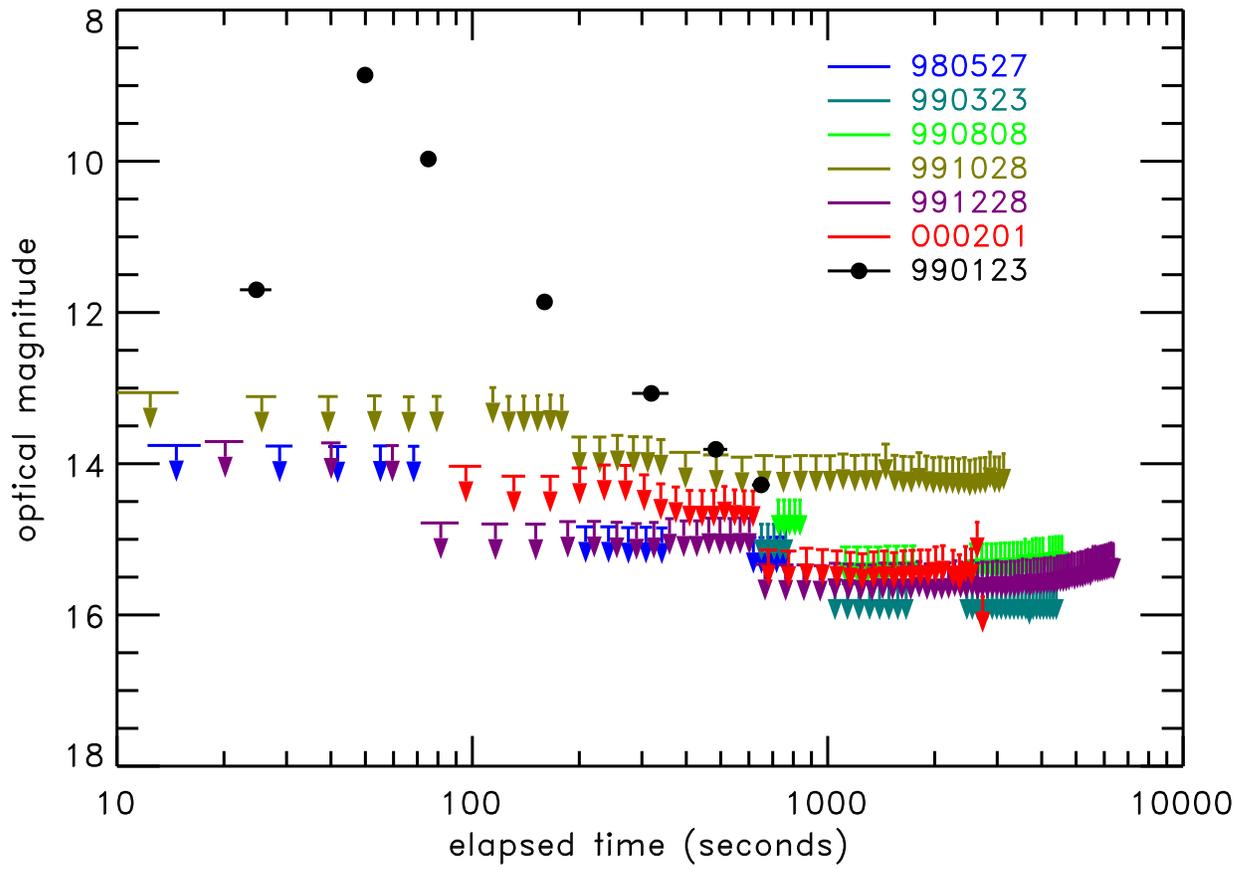
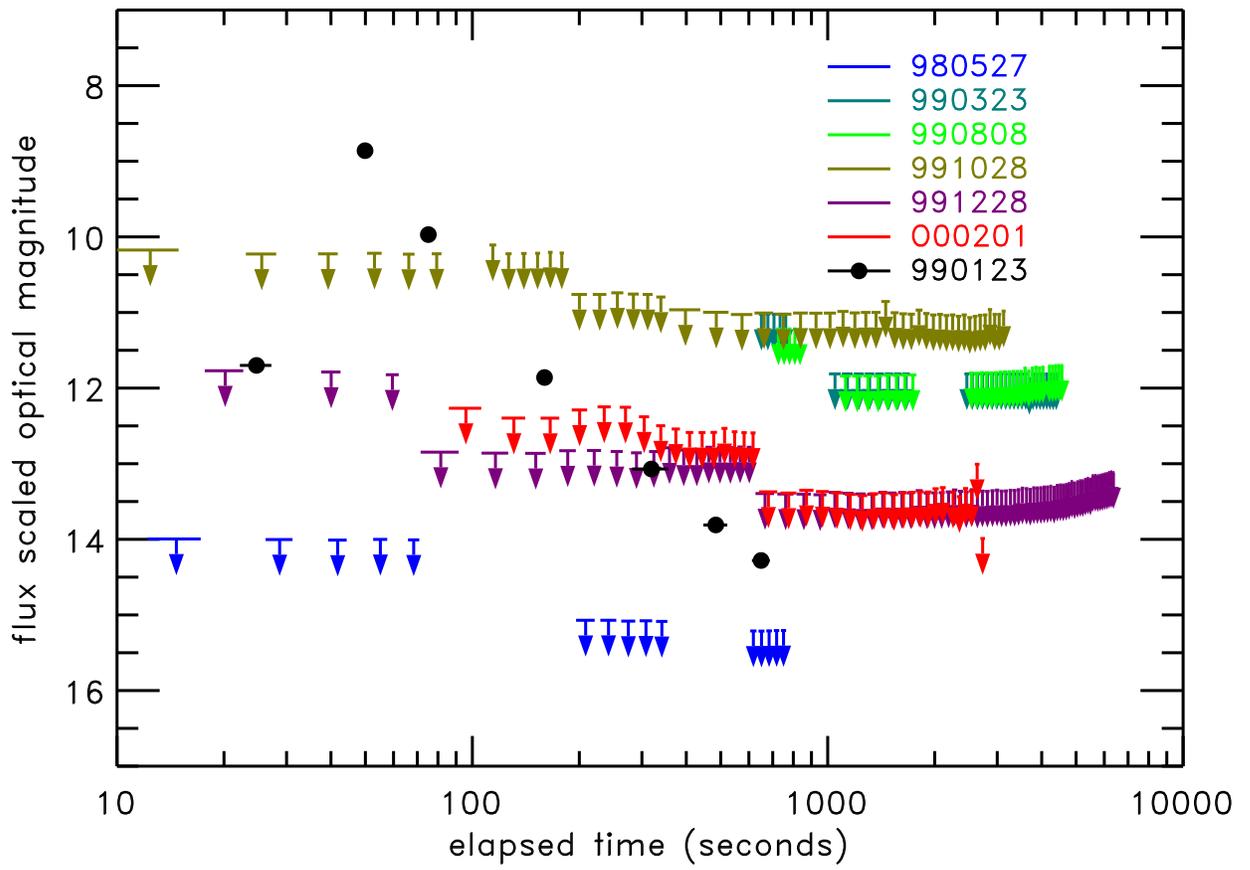
\end{document}